# "Zero resistance" when metals mixed with insulators

Ya-Dong Gu[1,2], Ji-Hai Yuan[1,2], Zhi-An Ren[1,2*]

[1] Institute of Physics and Beijing National Laboratory for Condensed Matter Physics, Chinese Academy of Sciences, Beijing 100190, China

[2] School of Physical Sciences, University of Chinese Academy of Sciences, Beijing 100049, China

[*] Corresponding author. E-mail: renzhian@iphy.ac.cn

**Abstract**

A false "zero resistance" behavior was observed during our study on the search of superconductivity in Ge-doped $GaNb_4Se_8$. This "zero resistance" was proved to be caused by open-circuit in multi-phase samples comprised of metals and insulators by measuring with four-probe method. The evidence strongly suggests that the reported superconductivity in hydrides should be carefully re-checked.

Superconductors exhibit two well-known effects: zero resistance and perfect diamagnetism. Both effects are independent and unique, and are simple and beautiful. In principle, either of the two effects can prove the existence of superconductivity, because no other state of matter exhibits these properties so far. Researchers routinely present both properties as complementary evidences to confirm the discovery of a new superconductor. But in some experiments, when researchers are dealing with multi-phase samples, highly air-sensitive powders, metastable samples, tiny nanotubes or nanowires, interfaces, high pressure samples with extreme conditions, *etc.*, the measurement and judgement of superconductivity becomes quite difficult and people often make mistakes even with the help of the most advanced modern instruments. Beyond measurements, certain conceptual confusions frequently arise in superconductivity research nowadays. These conceptual confusions include:

1. The zero-resistance criterion has been widely and incorrectly replaced by a resistive transition or drop behavior.

2. The shift of the resistive transition under magnetic fields has been widely and incorrectly used as a criterion for judging superconductivity.

3. The background subtraction method has been widely and incorrectly used for calculating the diamagnetic signal.

These erroneous criteria have not only caused significant confusions among beginners who are just entering this field, but have also misled experienced researchers and senior scientists, leading to widespread misconceptions for the judgment of a superconductor. In 2024, fifteen eminent scientists made a conclusion that "hydride superconductivity is real" based on unreliable data and incorrect criteria [1]. This raises a question that why the field of superconductivity has become so chaotic in recent years?

Here we present a false "zero resistance" phenomenon observed in Ge-doped $GaNb_4Se_8$ samples to explain that why the hydride superconductivity is still in need of critical scrutiny, or to put it bluntly, there's no solid evidence for the occurrence of superconductivity in these hydrides under high pressures.

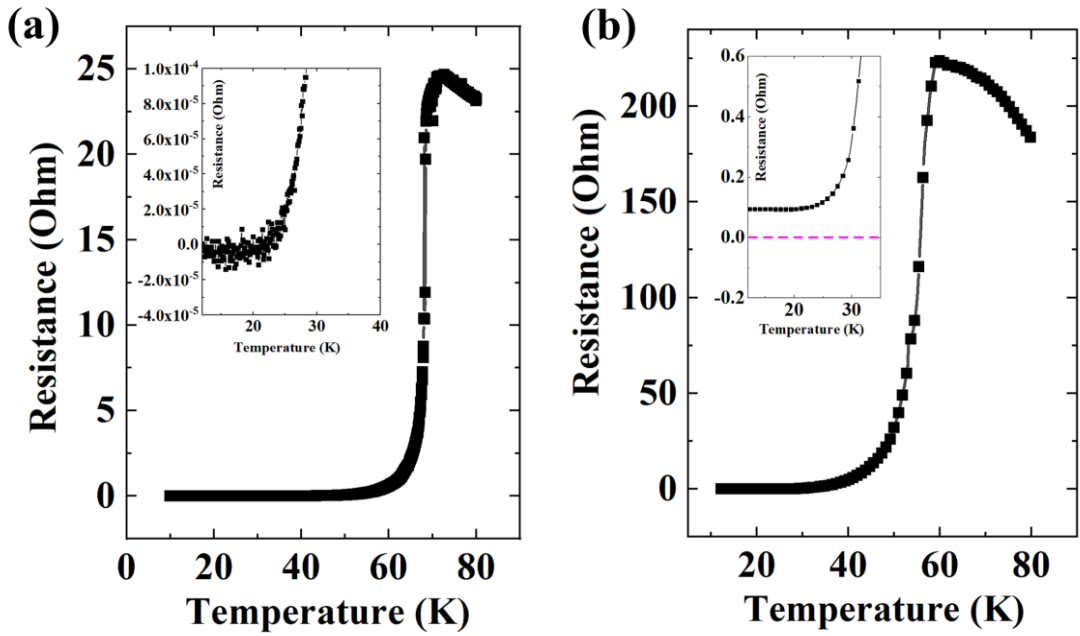

Figure 1. Temperature dependence of the electrical resistance for two Ge-doped GaNb$_4$Se$_8$ samples showing false "zero resistance" transitions, and a "zero resistance" is obtained in (a) and a positive residual resistance is shown in (b).

During our study on the superconductivity of Ge-doped GaNb$_4$Se$_8$ which mainly comprises the Mott insulating GaNb$_4$Se$_8$ and the metallic NbSe$_2$, we often observed another type of "zero resistance" besides our claimed superconducting zero resistance in Ref. [2]. The data are shown in Fig. 1 measured with the standard four-probe method with a 16 T PPMS. A sharp resistance transition at 72 K was observed in a sample exhibited in Fig. 1a, and the resistance actually drops to zero at lower temperatures (the resistance values are about $10^{-5}$ ohm corresponding to measured voltage about $10^{-8}$ V with the current of 1 mA). As more samples were measured, we found that the residual resistance becomes interesting in different samples, it might be positive, zero, or negative values, and usually a flat line vs. temperature but sometimes not. Fig. 1b shows another sample with a positive residual resistance. This made us thinking about the open-circuit of voltage probes, in which the values of the residual resistance depend on the electromagnetic environment of the electric circuit. But when we removed the sample for inspection, every probe was found to be normal. This "zero resistance" effect is stable and reproducible even after several months on the same sample.

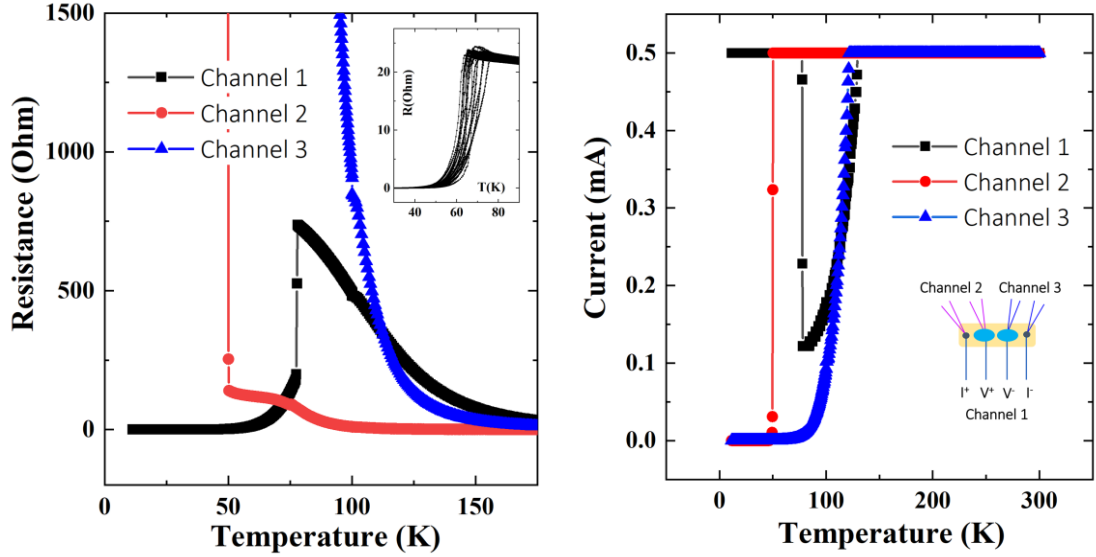

Figure 2. (a) Measurement of the temperature dependence of the electrical resistance for one Ge-doped $GaNb_4Se_8$ sample with three channels. Inset shows another sample's resistive transition shifting under magnetic fields. (b) The changes of the applied current in the three channels showing the open-circuit of two voltage probes at low temperatures. Inset shows the wire connections on the sample for three channels.

Soon we designed another simple measurement method to solve this problem. As shown in Fig. 2, we used three channels of the PPMS to measure one sample simultaneously. And we quickly found that at low temperatures, the conductive path between $I^+$ and $I^-$ probes kept unobstructed, but the voltage probes $V^+$ and $V^-$ were isolated at some temperatures and their currents become zero. The explanation is simple, because in the multi-phase Ge-doped $GaNb_4Se_8$ samples, the main component of the sample is $GaNb_4Se_8$, which became totally insulating at low temperatures and isolated the voltage probes $V^+$ and $V^-$. While sometimes the impurity $NbSe_2$ (which is metallic and has low resistivity) makes a conductive path between $I^+$ and $I^-$ probes. This makes the false "zero resistance" obtained by the equipment. This false "zero resistance effect" happens when metals mixed with insulators by measuring with four-probe method.

We note that this false "zero resistance transition" shifts with applied magnetic field or applied current clearly as shown in the inset of Fig. 2a for a field up to 16 T, but not in a monotonic manner, which will not be discussed here.

Now let's discuss about the hydride superconductivity. In Ref. [3], the authors presented four experimental evidences to claim superconductivity at 203 K in $H_3S$, which are:

1. "a sharp drop of the resistivity to zero"
2. "a decrease of the transition temperature with magnetic field"
3. "magnetic susceptibility measurements"
4. "a pronounced isotope shift of $T_c$"

Let's review these evidences from the last to the first. Some researchers may remember the debate of isotope effects in iron-based superconductors in an ISS-conference in some year, in which two contradictory experimental results were reported but both were theoretically explained. From an experimentalist's standpoint, these isotope effect experiments should not have been conducted except for wasting funds before the superconducting $T_c$ can be stably and identically reproduced again and again, which is extremely difficult in laboratory for these high-$T_c$ superconductors. The hydrides are encountering the same situation. A reproduction experiment for $H_3S$ by the high-pressure expert SHIMIZU shows the onset resistance transition from the temperature of 70 – 180 K in Ref. [4].

For the magnetic susceptibility measurements, back to 2006, when REN worked in AKIMITSU lab on the superconductivity of carbon nanotubes, after two years' study, REN and AKIMITSU reached the consensus that the background subtraction method had been widely misused to calculate resistance or susceptibility, and hence too many mistakes were made in that community. Basically, one can get desired curves with enough measurements and data selection plus background subtraction. Particularly for susceptibility measurements of weak signals, many ferromagnetic contamination signals have been subtracted into superconducting evidences mistakenly. The hydrides are also encountering the same background subtraction situation. The susceptibility data are unreliable.

For "a decrease of the transition temperature with magnetic field", which is actually a behavior of superconducting transitions, but not a criterion for judging superconductivity. As stated above, a false "resistance transition" shifts with applied magnetic field.

For the most important and solid evidence, "a sharp drop of the resistivity to zero", it is clear that the hydrides are multi-phases and include insulators as shown in Extended Data Figure 2 in Ref. [1], which just resembles the multi-phase $GaNb_4Se_8$ samples. The sharp drop is also clearly reproduced in the above $GaNb_4Se_8$ experiments by open-circuit, but there is still possibility that the hydride superconductivity is different and true. For those reported resistance transitions of hydrides, there are two most significant points of doubt. The first is the unusual width of the transition (as small as 0.5 K for a 203 K superconductor in Fig. 1d in arXiv Ref. [1]), for which HIRSCH has kept on opposing with reasonable judgments [5]. The second is the residual resistance at low temperatures, which can be observed as positive values in Extended Data Figure 3e in Ref. [1], Extended Data Fig. 3c in Ref. [6], and negative values in Fig. 1 and Extended Data Fig. 6c in Ref. [6]. And actually, these residual resistance values are quite large in above figures and cannot be neglected. These evidences strongly suggest that the measured resistance transitions of hydrides should be carefully re-checked.

In summary, we observed two types of "zero resistance transition" during our study on the search of superconductivity in Ge-doped $GaNb_4Se_8$. After careful examinations, we conclude that one of them is the superconducting phenomenon, and another type was proved to be caused by open-circuit in multi-phase samples comprised of metals and insulators by measuring with four-probe method.

**Acknowledgments**

This work was supported by the Strategic Priority Research Program of Chinese Academy of Sciences (Grant No. XDB25000000), the CAS Superconducting Research Project (Grant No. SCZX-0103), the National Key Research and Development Program of China (Grant No. 2021YFA1401800).